\documentclass[a4paper,11pt]{article}

\usepackage{cite}
\usepackage[cmex10]{amsmath}
\usepackage{array}
\usepackage{unit}
\usepackage{graphicx}
\usepackage{amssymb}
\usepackage{ascmac}

\begin{document}
\title{%
Estimation of Power System Inertia\\
Using Nonlinear Koopman Modes
}%
\author{%
Yoshihiko Susuki, Ryo Hamasaki, and Atsushi Ishigame\footnote{They are with Department of Electrical and Information Systems, Osaka Prefecture University, Japan, \texttt{susuki@eis.osakafu-u.ac.jp}, \texttt{susuki@ieee.org}.}
}%
\date{}

\maketitle

\begin{abstract}
We report a new approach to estimating power system inertia directly from time-series data on power system dynamics.  
The approach is based on the so-called Koopman Mode Decomposition (KMD) of such dynamic data, which is a nonlinear generalization of linear modal decomposition through spectral analysis of the Koopman operator for nonlinear dynamical systems.  
The KMD-based approach is thus applicable to dynamic data that evolve in nonlinear regime of power system characteristics.  
Its effectiveness is numerically evaluated with transient stability simulations of the IEEE New England test system.  
\end{abstract}

\section{Introduction}
\label{sec:intro}

Inertia is a key physical parameter of power systems that determines their frequency dynamics and stability. 
Due to the large penetration of renewables operating in grid-connected mode, it is of practical importance to estimate the system-wide inertia accurately for analysis and control of the system frequency; see e.g. \cite{Ulbig_IFACWC2014}. 
Also, the so-called synthetic inertia has been proposed to emulate the inertial responses of rotating generators with grid-connected inverters; see e.g. \cite{Fairley:2016,Hikihara_IEICETEA90}.   
Therefore, the system-wide inertia is expected to significantly vary along the day, and its estimation in real time is an important problem.

Many groups of researchers have developed methods for inertia estimation from data derived with practical measurements.  
In \cite{Inoue:1997} the authors proposed to use measured frequency transients for estimation of power system inertia constant by polynomial approximation of such transient data.  
In \cite{Jones:2004} the author used the ARIMAX model for power system parameters including the inertia constant.  
In \cite{Ota:2007} the authors used synchrophasor measurements for parameter estimation of an interconnected portion of the Japanese power system.  
In \cite{Wall:2014} the authors proposed an online algorithm for estimating the time of disturbance and inertia parameter after a disturbance.  
In \cite{Guo:2014} the authors proposed to use the weighted least-squares method for adaptive parameter estimation of power system dynamic models from data obtained from wide area measurement systems.  
Also, in \cite{Ashton:2015} the authors used synchrophasor measurements in the Great Britain power system for its inertia estimation.  

In this paper, we introduce an alternative approach to the inertia estimation directly from time-series data on power system dynamics.  
The main idea of this paper is to utilize the Koopman Mode Decomposition (KMD) for the inertia estimation directly from PMU measurements or numerical data.  
KMD is a relatively new technique of nonlinear time-series analysis in a non-parametric manner \cite{Mezic_ND41,Rowley_JFM641}.  
This is based on spectral properties of the Koopman operator for a underling nonlinear dynamical model, which corresponds to state-space models of power systems, and has been used in power systems engineering; please refer to \cite{Susuki_IEEETPWRS26,Susuki_IEEETPWRS27,Susuki_IEEETPWRS29,Raak_IEEETPWRS31,Susuki_NOLTAIEICE7,Barocio_IEEETPWRS30,Dubey_EPES89}.   
An important point here is that via the Koopman operator, linear techniques can be used for analysis and control of nonlinear power system dynamics under the mathematically-rigor support.  
The contribution of this paper is to develop a methodology for inertia estimation directly from data of possibly nonlinear (large) excursions in rotor speeds and to establish its effectiveness through transient stability simulations on the IEEE New England test system \cite{Pai:1989}.  
This wad made possible by introducing the KMD into the inertia estimation problem, which is the novelty of this paper.   

The rest of the paper is organized as follows.  
The theory and computations of KMD are summarized in Section~\ref{sec:KMD}.  
The main idea of this paper is introduced in Section~\ref{sec:main}.  
Numerical investigations of the main idea are conducted in Section~\ref{sec:numerics}.  
Concluding remarks are made in Section~\ref{sec:outro} with ongoing and future works.

\section{Nonlinear Koopman Modes}
\label{sec:KMD}

In this section, we introduce the mathematical idea of KMD (Koopman Mode Decomposition) based on \cite{Susuki_NOLTAIEICE7} in the context of power system dynamics.  
In the following, we suppose that the power system dynamics are represented well by the following ordinary differential equation on a finite-dimensional smooth manifold $\mathbb{X}$: for continuous time $t\in\mathbb{R}$, 
\begin{equation}
\frac{\dd}{\dd t}\vct{x}(t)=\vct{F}(\vct{x}(t)),
\label{eqn:model}
\end{equation}
where $\vct{x}\in\mathbb{X}$ is the state of the power system model including rotor angles and rotor speeds, and $\vct{F}: \mathbb{X}\to\mathrm{T}\mathbb{X}$ (tangent bundle of $\mathbb{X}$) a 
nonlinear function representing the nonlinear electro-mechanical characteristics of rotating generators, flux decays inside such generators, control mechanisms, and so on.  
The function $\vct{F}$ is assumed to be tractable in the region we are interested.  
For this equation, the following finite time-$t$ map $\vct{S}_t$ is defined as
\[
\vct{S}_t: 
\mathbb{X}\to\mathbb{X};~
\vct{x}(0)\mapsto\vct{x}(t)=\vct{x}(0)+\int^t_0\vct{F}(\vct{x}(\tau))\dd\tau.
\]
The one-parameter group of maps, $\{\vct{S}_t; t\in\mathbb{R}\}$, is called the \emph{flow}.  

Here we introduce the Koopman operators for the flow.  
To do this, the so-called \emph{observable} $f$ is introduced as a scalar-valued function defined on $\mathbb{X}$, namely
\[
f: \mathbb{X}\to\mathbb{C}.
\]
This observable is a mathematical formulation of observation or measurement in a power system such as voltage phasors and power flows.  
Below, we will denote by $\mathcal{F}$ a given space of observables.  
For an observable $f\in\mathcal{F}$, the \emph{Koopman operator} $\mathbf{U}_t$ for (\ref{eqn:model}) maps $f$ into a new function as follows:
\[
\mathbf{U}_tf:=f\circ\vct{S}_t.
\]
That is, the Koopman operator $\mathbf{U}_t$ describes the time $t$ evolution of observation (or measurement) along the state's dynamics; in fact we can write the time evolution of the observation, $y(t):=f(\vct{x}(t))$, as follows:
\[
y(t)=f(\vct{x}(t))=f(\vct{S}_t(\vct{x}(0))
=(\mathbf{U}_tf)(\vct{x}(0)).
\]
Although the model (\ref{eqn:model}) can be nonlinear and evolve in the finite-dimensional space, the Koopman operator is \emph{linear} but \emph{infinite-dimensional}.  
This type of composition operator is defined for a large class of nonlinear dynamical systems \cite{Lasota:1994} and does not rely on linearization; indeed, it captures the full information on the original nonlinear system (\ref{eqn:model}) including stationary and transient behaviors.  

In the developing theory reviewed in \cite{Marko_CHAOS22,Mezic_ARFM45,Susuki_NOLTAIEICE7}, spectral properties of the Koopman operator are of paramount importance.    
The pair of eigenvalue $\lambda\in\mathbb{C}$ and eigenfunction $\phi_\lambda\in\mathcal{F}\setminus\{0\}$ of the Koopman operator $\mathbf{U}_t$ is defined as follows:
\[
\mathbf{U}_t\phi_\lambda=\exp(\lambda t)\phi_\lambda.
\]
Generally, the number of pairs of eigenvalues and eigenfunctions of the Koopman operator is not finite.  
Here, for general situations of synchronized measurements at multiple locations in a power system, we consider a vector-valued observable $\vct{f}:=(f_1,\ldots,f_m)^\top: \mathbb{X}\to\mathbb{C}^m$ ($f_i\in\mathcal{F}$; $\top$ denotes the transpose operation of vectors and matrices), and we assume that each $f_i$ is expanded in terms of the eigenfunctions of the Koopman operator:
\[
\vct{f}=\sum^\infty_{j=1}\phi_{\lambda_j}\vct{V}_j
\]
where $\vct{V}_j\in\mathbb{C}^m$ is the coefficient vector for expansion.  
Then, the time evolution of observation, $\vct{y}(t):=\vct{f}(\vct{x}(t))$, along the state's dynamics is derived in \cite{Mezic_ARFM45} as follows:
\begin{equation}
\vct{y}(t)
=\sum^\infty_{j=1}\exp(\lambda_jt)\phi_{\lambda_j}(\vct{x}(0))\vct{V}_j.
\label{eqn:KMDc}
\end{equation}
This type of time-series decomposition based on the Koopman eigenvalues and eigenfunctions is named in \cite{Rowley_JFM641} as the \emph{Koopman Mode Decomposition} (KMD). 
Here, under equi-sampling ($T$) of the observation, we have the discrete version of KMD: for $k=0,1,2,\ldots$
\[
\vct{y}_k:=\vct{y}(kT)
=\sum^\infty_{j=1}\tilde{\lambda}_j^k\vct{\tilde{V}}_j,
\]
with
\[
\tilde{\lambda}_j:=\exp(\lambda_jT),~~~
\vct{\tilde{V}}_j:=\phi_{\lambda_j}(\vct{x}(0))\vct{V}_j.
\]
This formulation is suitable to data analytics, and thus many algorithms of computing a finite number of pairs $(\tilde{\lambda}_j,\vct{\tilde{V}}_j)$ directly from finite samples of time-series data (observation) $\{\vct{y}_k\}$ have been developed: see \cite{Marko_CHAOS22,Susuki_NOLTAIEICE7} and references therein.  
Especially, it is shown in \cite{Susuki_CDC15} that the Prony analysis, which has been widely used in power system analysis (see e.g. \cite{Hauer_IEEETPS5,Nabavi_ACC14}), provides a finite approximation of KMD.  
Its rigor mathematical proof for ergodic dynamical systems is presented in \cite{Arbavi_Preprint:2017}.   
The multichannel version of the Prony analysis, which is called in \cite{Susuki_CDC15} the vector Prony analysis, is used in Section\,\ref{sec:numerics}.

\section{Main Idea}
\label{sec:main}

In this section, we present the main idea of the present paper: to apply the KMD to the inertia estimation problem.  
Now consider a power system of $N$ rotating generators with non-zero inertia.  
The electro-mechanical dynamics are represented by the so-called swing equations that can be included in (\ref{eqn:model}).  
The equations are given in \cite{Kundur_PSSC} as
\begin{equation}
\left.
\begin{array}{rcl}
\displaystyle\frac{\dd\delta_i}{\dd t} &=& \omega_i
\\\noalign{\vskip 2mm}
\displaystyle M_i\frac{\dd\omega_i}{\dd t} &=& \Delta P_i
\end{array}
\right\}
\label{eqn:SwingSingle}
\end{equation}
where $i\in\{1,\ldots,N\}$ is the integer index of generators.  
The variable $\delta_i$ is the rotor angle of generator \#$i$ relative to a synchronously-rotating axis, and $\omega_i$ its relative rotor speed, and $\Delta P_i$ its accelerating power.  
The parameter $M_i$ denotes the inertial parameter of the $i$-th generator, which is our target of the estimation problem.  
An essential point for the following inertia estimation is that the two quantities $\omega_i$ and $\Delta P_i$ are scalar-valued functions of the states of the underlying model (\ref{eqn:model}) including $\delta_i$ and $\omega_i$, that is, \emph{observables} of the model.  
Here, from (\ref{eqn:SwingSingle}) we have
\begin{equation}
[M_1,M_2,\ldots,M_N]\frac{\dd}{\dd t}\vct{\omega}(t)=\Delta P(t) 
\label{eqn:SwingSystem}
\end{equation}
where $\vct{\omega}(t):=[\omega_1(t),\ldots,\omega_{N}(t)]^\top$, and $\Delta P(t):=\sum^N_{i=1}\Delta P_i(t)$ denotes the net accelerating power of the system, which corresponds to the net exchange power of the system to its outside. 
Now, as an initial study, we suppose that (possibly) nonlinear time-evolutions of $\vct{\omega}(t)$ and $\Delta P(t)$ are available by practical measurements or simulations, and that they contain only distinct components (peaks) in the frequency spectra.  
Here, since the two quantities $\omega_i$ and $\Delta P_i$ are observables of the underlying model, as in (\ref{eqn:KMDc}) we can decompose their time-evolutions in terms of eigenvalues and modes of the Koopman operator of the underlying model (\ref{eqn:model}) as follows:
\begin{equation}
\left[
\begin{array}{c}
\vct{\omega}(t) \\ \Delta P(t)
\end{array}
\right]=
\sum^\infty_{j=1}\exp(\lambda_jt)
\left[
\begin{array}{c}
\vct{V}_j^\omega \\\noalign{\vskip 1mm} V_j^P
\end{array}
\right]
\label{eqn:KMD}
\end{equation}
where $\lambda_j\in\mathbb{C}$ denotes the $j$-th eigenvalue of the Koopman operator and $[(\vct{V}_j^\omega)^\top V_j^P]^\top\in\mathbb{C}^{N+1}$ the corresponding Koopman mode.  
This decomposition is conducted directly from sampled time-series without any use of models as shown in the previous section and in \cite{Susuki_NOLTAIEICE7}.  
The important point here is that both the time-evolutions of $\vct{\omega}(t)$ and $\Delta P(t)$ are decomposed with a common set of Koopman eigenvalues (eigenfrequencies) even in the nonlinear regime of state dynamics.   
By substituting (\ref{eqn:KMD}) into (\ref{eqn:SwingSystem}), we have
\begin{equation}
[M_1,M_2,\ldots,M_N]
\sum^\infty_{j=1}\lambda_j\exp(\lambda_jt)
\vct{V}_j^\omega \nonumber\\
=
\sum^\infty_{j=1}\exp(\lambda_jt)V_j^P.
\label{eqn:hoge}
\end{equation}
Assume that the $\lambda_j$ are distinct.  
Then, since $t$ is arbitrary, we have
\[
\underset{\bf H}{
\underbrace{
\left[
\begin{array}{c}
\lambda_1(\vct{V}^\omega_1)^\top \\ \lambda_2(\vct{V}^\omega_2)^\top \\ \vdots
\end{array}
\right]
}}
~
\underset{M}{
\underbrace{
\left[
\begin{array}{c}
M_1 \\ M_2 \\ \vdots \\ M_N
\end{array}
\right]
}}
=
\underset{\bf b}{
\underbrace{
\left[
\begin{array}{c}
V^P_1 \\ V^P_2 \\ \vdots
\end{array}
\right]
}}.
\]
Here, although $\mathbf{H}$ and $\mathbf{b}$ are now infinite-dimensional, in numerics, as seen in \cite{Susuki_NOLTAIEICE7} we can obtain their finite truncations like
\begin{equation}
\underset{{\sf H}_m}{
\underbrace{
\left[
\begin{array}{c}
\lambda_1(\vct{V}^\omega_1)^\top \\ \lambda_2(\vct{V}^\omega_2)^\top \\ \vdots 
\\ \lambda_m(\vct{V}^\omega_m)^\top
\end{array}
\right]
}}
~
\underset{M}{
\underbrace{
\left[
\begin{array}{c}
M_1 \\ M_2 \\ \vdots \\ M_N
\end{array}
\right]
}}
=
\underset{b_m}{
\underbrace{
\left[
\begin{array}{c}
V^P_1 \\ V^P_2 \\ \vdots \\ V^P_m
\end{array}
\right]
}}.
\label{eqn:truncated}
\end{equation}
Thus, if the finite truncated $\mathsf{H}_m$ is of full-column rank, then the unique solution $(\mathsf{H}^\top_m\mathsf{H}_m)^{-1}\mathsf{H}^\top_m\vct{b}_m$ is obtained as the estimated inertial parameters $\vct{M}$. 
If this is not the case, then we will use pseudo-inverse of ${\sf H}_m$ for obtaining an estimated vector of $\vct{M}$. 
The inertial parameter of the power system corresponds to $\displaystyle \sum^N_{i=1}M_i$, which we call system-wide inertia in the following investigations.  

It should be noted that the KMD-based inertia estimation in (\ref{eqn:truncated}) is data-driven and model-free.  
No limitation of its application exits if sufficient dynamic data are available.  
Also, by exploiting the KMD, in (\ref{eqn:hoge}) we have avoided numerical differentiation of the rotor speeds $\omega_i(t)$.  
This takes advantage to handle dynamic data with additive noise, which are common in the PMU application.  
In addition to this, different from \cite{Inoue:1997} using finite degree of polynomial, we do not make approximation of transient swings; in theory, by taking the \emph{infinite} sum of exponentials, it is possible to derive a complete representation of the transient swings even if they evolve in nonlinear regime of power system characteristics.  
Therefore, we expect that the KMD-based inertia estimation is generally applicable and works well for a wide class of observational data in power systems.

\section{Numerical Investigations}
\label{sec:numerics}

\begin{figure}[t]
\centering
\includegraphics[width=.65\textwidth]{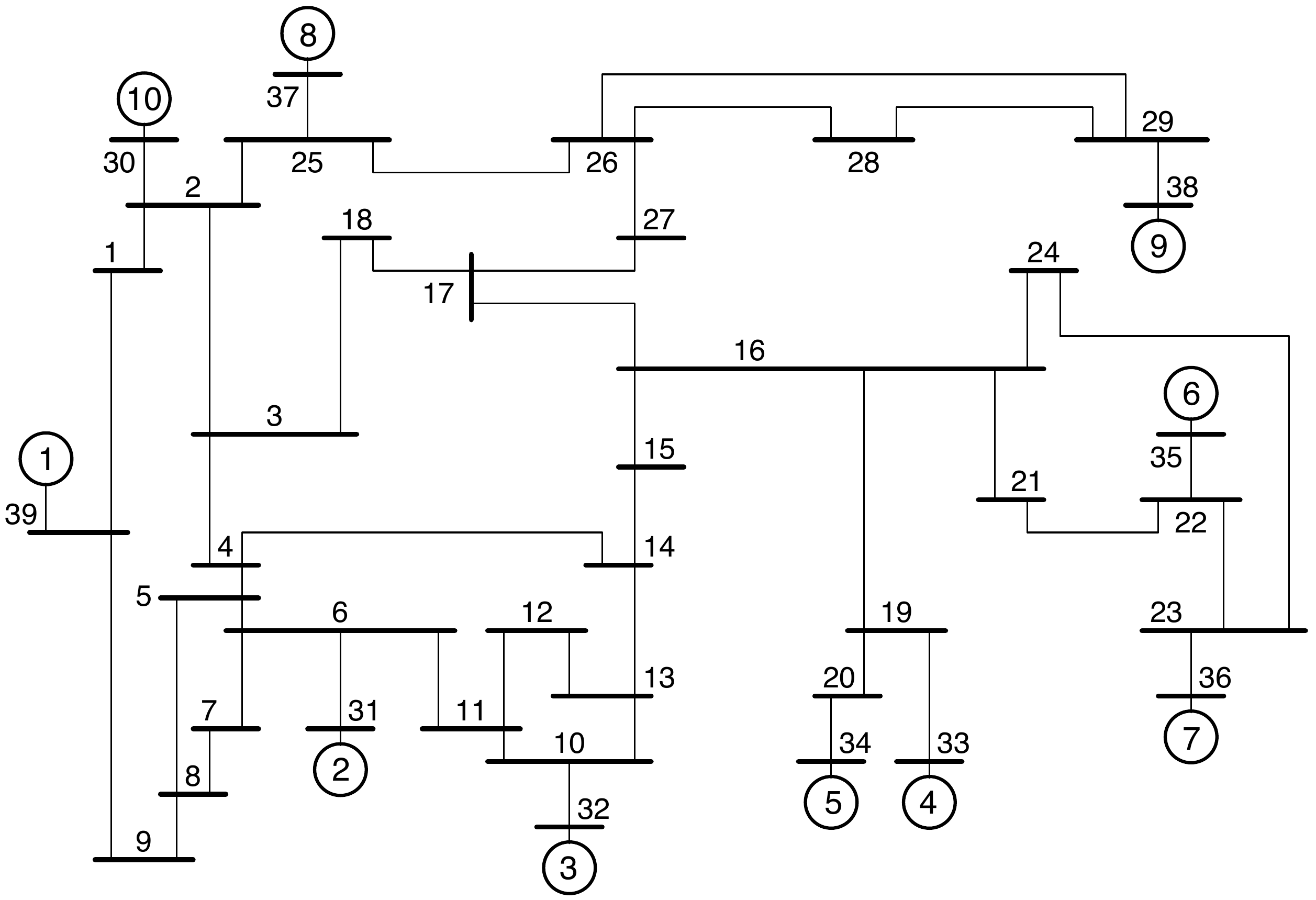}
\caption{%
One-line diagram of IEEE New England test system. 
The 9 generators, \#2 to \#10, exhibit transient dynamics that we address for the inertia estimation.  
}%
\label{fig:NE}
\end{figure}

In this section, we present a series of numerical investigations of the inertia estimation based on KMD.  
The well-known IEEE New England test system is used for the investigations.  
Its one-line diagram is shown in Figure\,\ref{fig:NE} and has the 10 generators (\#1 to \#10).  
Here we have assumed that generator \#1 is the infinite bus with no dynamics.  
The inertial parameters are presented in \cite{Pai:1989} as shown in Table\,\ref{tab:true}.  
The system-wide inertia here corresponds to the sum of inertia of the 9 generators (\#2 to \#10), which is the target parameter of the estimation problem.

\begin{table}[t]
\centering
\caption{Inertial Parameters in IEEE New England Test System}
\label{tab:true}
\begin{tabular}{ccr}\hline\hline\noalign{\vskip 1mm}
Generator \# & Inertial Parameter $M_i$ [p.u.] \\\hline
1 & --- & (infinite bus) \\
2 & 0.1607 \\
3 & 0.1899 \\
4 & 0.1517 \\
5 & 0.1379 \\
6 & 0.1846 \\
7 & 0.1401 \\
8 & 0.1289 \\
9 & 0.1830 \\
10 & 0.2228 \\
System-wide & 1.4998 & $\sum^{10}_{i=2}M_i$\\\noalign{\vskip 1mm}\hline\hline
\end{tabular}
\end{table}

\begin{figure}[t]
\centering
\includegraphics[width=.65\textwidth]{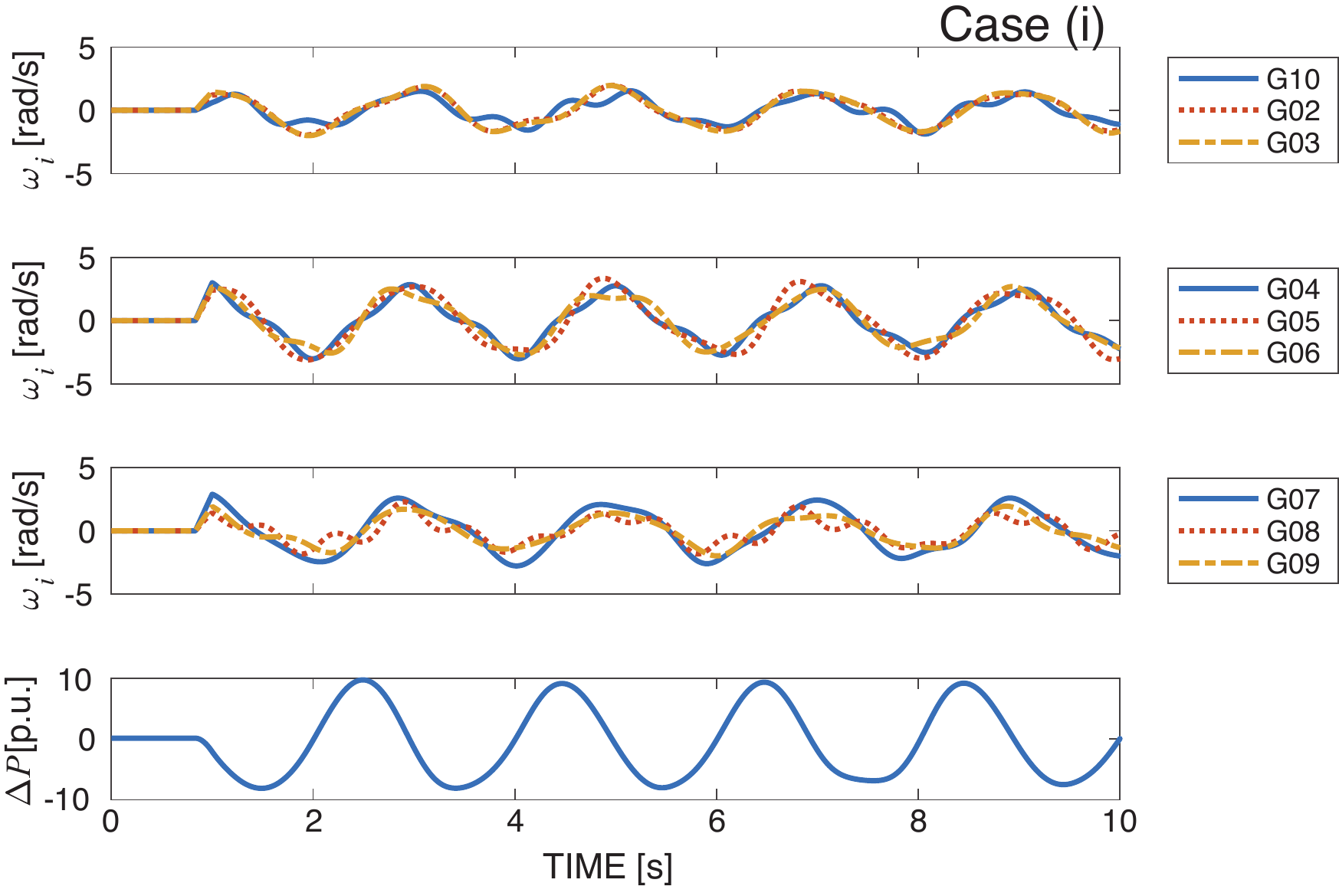}
\includegraphics[width=.65\textwidth]{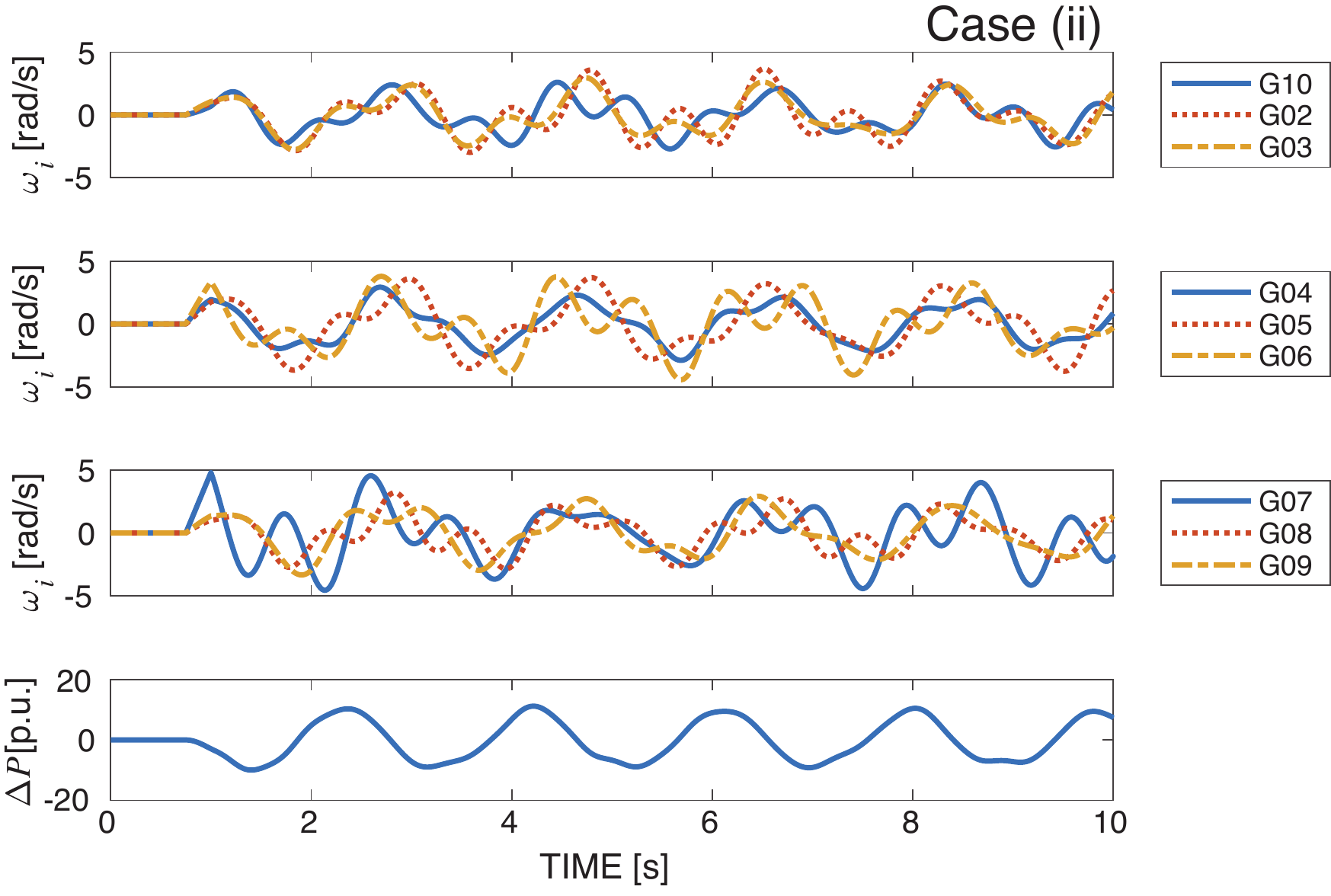}
\caption{%
Transient stability simulations of IEEE New England test system.  
The dynamic data are used for estimating the system-wide inertia via KMD.  
}%
\label{fig:dyn}
\end{figure}

For transient stability simulations we consider two cases of three-phase fault as follows:
\begin{itemize}
\item[(i)] The fault occurs near bus \#16 and line 16-17 is tripped. 
Its duration is 10 cycles of a 60-Hz sine wave (namely 1/6\,sec.); 
\item[(ii)] The fault occurs near bus \#23 and line 22-23 is tripped. 
Its duration is 15 cycles of a 60-Hz sine wave (namely 1/4\,sec.).  
\end{itemize}
The transient dynamics after these disturbances are simulated with the classical swing equations and are shown in Figure\,\ref{fig:dyn}.  
In Case (i), we see a global oscillatory mode in the relative rotor speeds $\omega_i(t)$ over the system, in which all the generators swing in a coherent manner.   
This type of the global mode appears in \cite{Susuki_JNLS09} and is a main cause of the nonlinear instability phenomenon in which the three different oscillatory modes---local-plant, inter-machine, and inter-area---interact to destabilize a power system.  
In this simulation, the net accelerating power $\Delta P(t)$ is likely a pure sine-wave.  
In Case (ii), we see complicated swings in $\omega_i(t)$, in particular, generator \#6 and \#7 exhibit large magnitudes compared with the others.  
This suggests that a local-plant mode mainly develops in the simulation.  
Here, the net accelerating power $\Delta P(t)$ is relatively distorted, where coupled swing dynamics of the 9 generators occur due to a nonlinear power coupling in the classical swing equations.  

\begin{figure}[t]
\centering
\includegraphics[width=.49\textwidth]{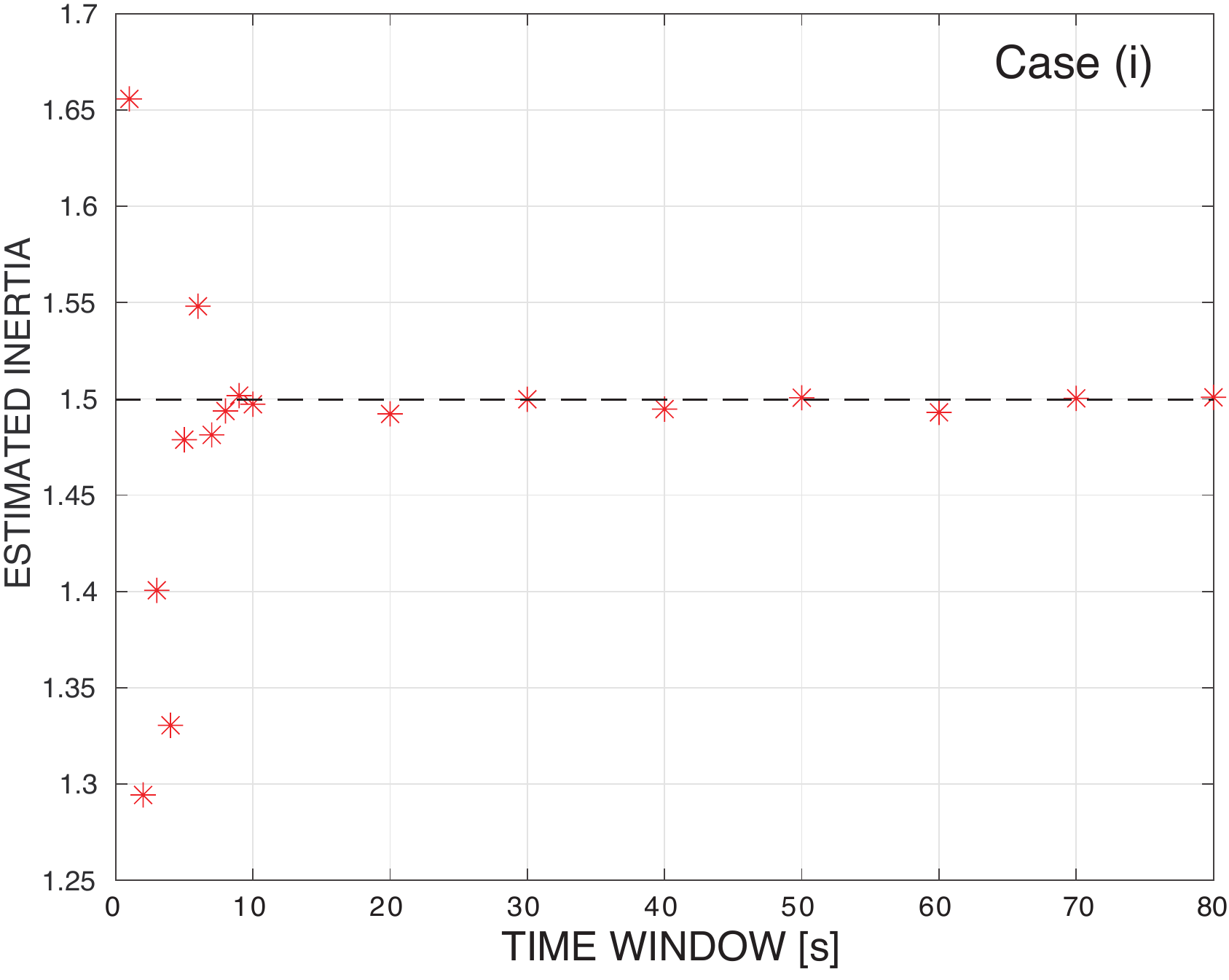}
\includegraphics[width=.49\textwidth]{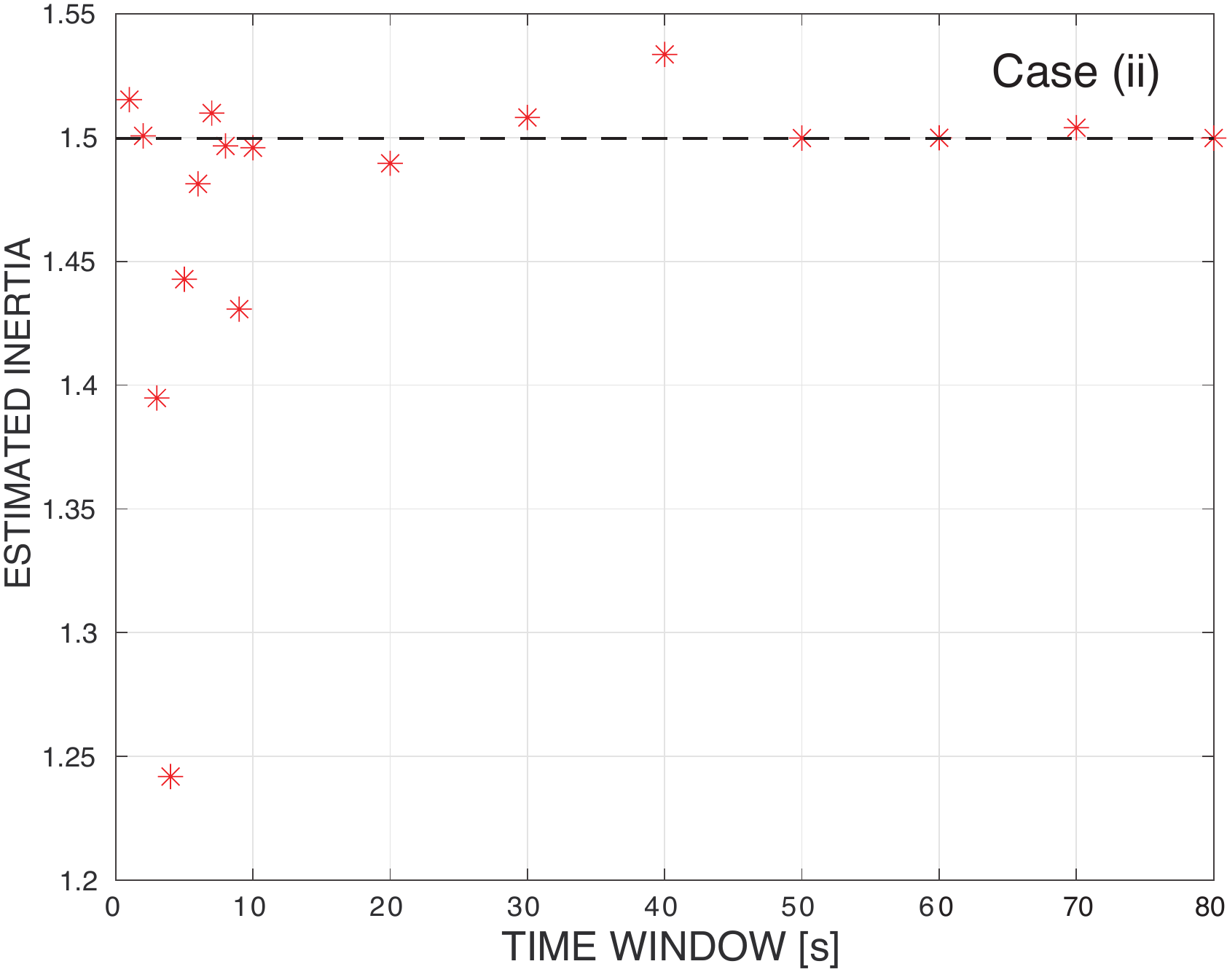}
\caption{%
KMD-based estimation of system-wide inertia for IEEE New England test system (i).  
The true value of inertia is 1.4998 and denoted by the horizontal broken line.  
}
\label{fig:estimation1}
\end{figure}

The results on estimation of system-wide inertia based on KMD are ploted in Figure\,\ref{fig:estimation1}.  
The KMD here was conducted for finite samples of time-series $\omega_i(t)$ and $\Delta P(t)$ under equi-sampling with $1/(60\U{Hz})$.  
The plots show how the choice of ``time window (of data)" affects the accuracy of inertia estimation.  
The term ``time window" implies the duration from the onset of fault clearing which we use for KMD.  
Therefore, the figures suggest that after ten seconds of analysis time (i.e. time constant for rotor angle stability as shown in Figure~\ref{fig:dyn}), the estimated values of system-wide inertia are consistent and close to the true value (1.4998).  

\begin{figure}[t]
\centering
\includegraphics[width=.49\textwidth]{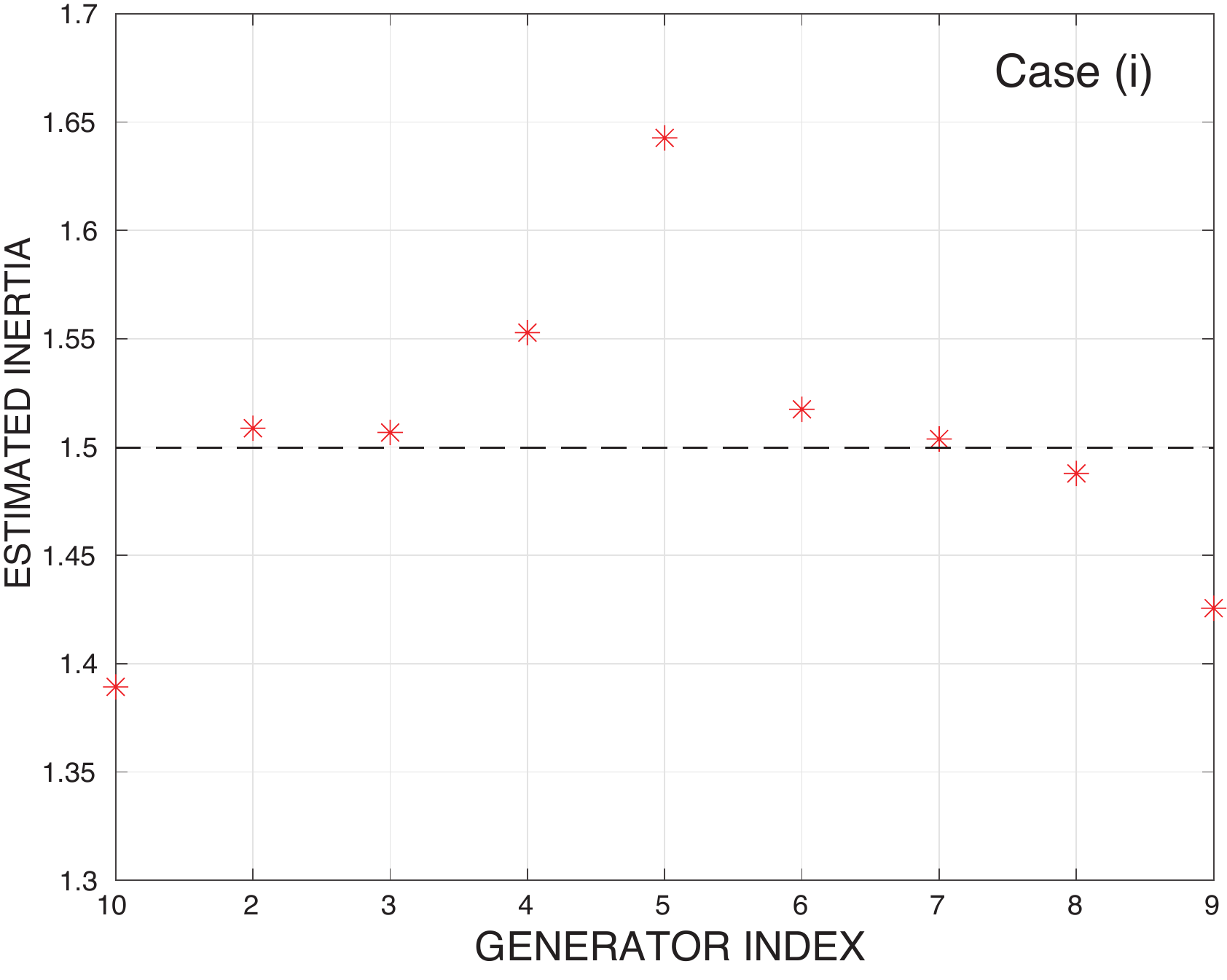}
\includegraphics[width=.49\textwidth]{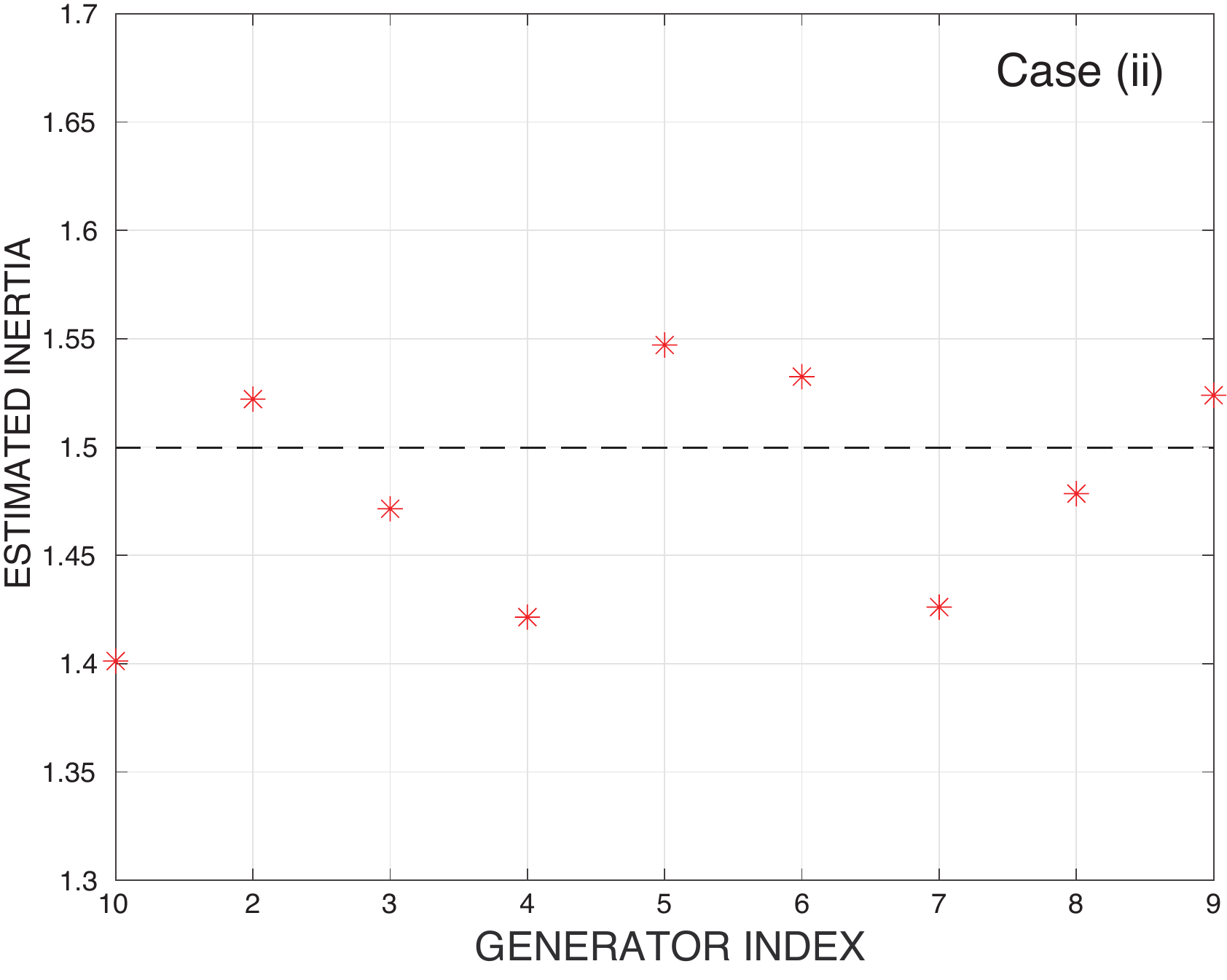}
\caption{%
KMD-based estimation of system-wide inertia for IEEE New England test system (ii). 
The true value of inertia is 1.4998 and denoted by the horizontal broken line.  
The horizontal axis denotes the index of generator for which dynamic data are not available for the estimation.  
}
\label{fig:estimation2}
\end{figure}

%
Above, we have used the dynamic data for all the 9 generators in the New England system.  
Here we investigate a case in which all the dynamic data are not available for the inertial estimation.  
For simplicity of analysis, we suppose that dynamic data on relative rotor speed from one generator are not available.  
For example, if $\omega_{10}(t)$ for generator \#10 is not available, then we conduct the estimation in (\ref{eqn:truncated}) for the vector $[M_2,M_3,\ldots,M_9]^\top$ using the time-series data $\{[\omega_2(t),\omega_3(t),\ldots,\omega_9(t),\Delta P(t)]^\top\}$.  
The results on estimation of system-wide inertia are shown in Figure\,\ref{fig:estimation2}.  
The value of time window is fixed at 10\,sec.  
In the figures, the horizontal axis denotes the index of generator for which dynamic data are not available for the estimation.  
The estimated values are close to the true one, 1.4998.  
Thus, we suggest that the KMD-based approach works even in partial information on relative rotor speeds.

\section{Concluding Remarks}
\label{sec:outro}

A new approach to the inertia estimation of power systems was presented in this paper.  
The key idea is to utilize the KMD (Koopman Mode Decomposition) that is a relatively new technique of nonlinear time-series analysis, which has a rigor mathematical background in Koopman operator theory of nonlinear dynamical systems.   
In this sense, the new approach works well for dynamic data evolving in nonlinear regime of power system characteristics.  
Numerical investigations in the IEEE New England test system indeed
show that the KMD-based approach works well for accurate estimation of inertia for nonlinear responses of rotor speeds and net accelerating power.  

The present paper is a short announcement of our work.  
Several follow-up studies to the work are needed and ongoing.  
One is to assess how adding noise to dynamic data affects the accuracy of inertia estimation.  
Also, it is interesting to apply the new approach to data measured in practice with wide-area monitoring systems based on PMU.  
In addition to this, it is of practical importance to use the approach for estimating (possibly, time-varying) inertia of a power system with high penetration of renewable energy resources operating in grid-connected mode.

\section*{Acknowledgements}

The authors are grateful to Professor Yasunori Mitani, Professor Masayuki Watanabe, and Mr. Jun Terashi (Kyushu Institute of Technology) for their valuable discussions, and Mr.~Marcos Netto (Virginia Tech) for his careful reading of the manuscript.  
The work presented here is supported in part by JSPS KAKEN \#15H03964 and The Specific Support Project of Osaka Prefecture University.  

\vspace*{5mm}




%

%

\bibliographystyle{IEEEtran}
\bibliography{/Users/susuki/text/mybib/IEEEabrv,/Users/susuki/text/mybib/mezic,/Users/susuki/text/mybib/main-SSK,/Users/susuki/text/mybib/koopman,/Users/susuki/text/mybib/ds,/Users/susuki/text/mybib/power,/Users/susuki/text/mybib/ml}

\end{document}